\documentclass[lettersize,journal]{IEEEtran}
\usepackage{amsmath,amsfonts}
\usepackage{algorithmic}
\usepackage{algorithm}
\usepackage{array}
\usepackage[caption=false,font=normalsize,labelfont=sf,textfont=sf]{subfig}
\usepackage{textcomp}
\usepackage{stfloats}
\usepackage{url}
\usepackage{verbatim}
\usepackage{graphicx}
\usepackage{cite}
\usepackage{xcolor}
\usepackage{booktabs}
\usepackage{multirow}
\usepackage[table,xcdraw]{xcolor}

\begin{document}

\title{AuthenLoRA: Entangling Stylization with Imperceptible Watermarks for Copyright-Secure LoRA Adapters}

\author{Fangming Shi, Li Li, Kejiang Chen, Guorui Feng, Xinpeng Zhang
\thanks{This work was supported in part by the National Natural Science Foundation of China under Grant No. 62450067 and No. 62302286 and Research Fund of Key Lab of Education Blockchain and Intelligent Technology, Ministry of Education under Grant No. EBME25-F-04. \textit{Corresponding author: Xinpeng Zhang})}
\thanks{Fangming Shi, Li Li and Guorui Feng are with the School of Communication and Information Engineering, Shanghai University, Shanghai 200444, China (e-mail: shifangming0823@shu.edu.cn, llichn@shu.edu.cn, grfeng@shu.edu.cn)}
\thanks{Kejiang Chen are with the Department of Electronic Engineering and Information Science, University of Science and Technology of China, Hefei 230052, China (e-mail: chenkj@ustc.edu.cn)}
\thanks{Xinpeng Zhang is with the School of Communication and Information Engineering, Shanghai University, Shanghai 200444, China, and also with the School of Computer Science, Fudan University, Shanghai 200433, China (e-mail: xzhang@shu.edu.cn)}}

\markboth{SUBMISSION TO IEEE TRANSACTIONS ON MULTIMEDIA}
{Shell \MakeLowercase{\textit{et al.}}: A Sample Article Using IEEEtran.cls for IEEE Journals}
\IEEEpubid{0000--0000/00\$00.00~\copyright~2021 IEEE}

\maketitle

\begin{abstract}
Low-Rank Adaptation (LoRA) offers an efficient paradigm for customizing diffusion models, but its ease of redistribution raises concerns over unauthorized use and the generation of untraceable content. Existing watermarking techniques either target base models or verify LoRA modules themselves, yet they fail to propagate watermarks to generated images, leaving a critical gap in traceability. Moreover, traceability watermarking designed for base models is not tightly coupled with stylization and often introduces visual degradation or high false-positive detection rates.
To address these limitations, we propose AuthenLoRA, a unified watermarking framework that embeds imperceptible, traceable watermarks directly into the LoRA training process while preserving stylization quality. AuthenLoRA employs a dual-objective optimization strategy that jointly learns the target style distribution and the watermark-induced distribution shift, ensuring that any image generated with the watermarked LoRA reliably carries the watermark. We further design an expanded LoRA architecture for enhanced multi-scale adaptation and introduce a zero-message regularization mechanism that substantially reduces false positives during watermark verification.
Extensive experiments demonstrate that AuthenLoRA achieves high-fidelity stylization, robust watermark propagation, and significantly lower false-positive rates compared with existing approaches. Open-source implementation is available at: \url{https://github.com/ShiFangming0823/AuthenLoRA}.
\end{abstract}

\begin{IEEEkeywords}
LoRA Watermark, Copyright Protection, Text-to-Image Model.
\end{IEEEkeywords}

\section{Introduction}
Recent advances in diffusion models~\cite{DDPM, ImprovedDDPM, DDIM, StableDiffusionRepository, StableDiffusionv2GithubRepository} have significantly propelled text-to-image generation and exhibited impressive fidelity and semantic alignment with complex prompts. While demand grows for customized text-to-image generation, such as designers requiring brand-specific styles, users seeking self-representation in generated media, and unification of stylization in film and game art, full fine-tuning approaches incur prohibitively high training costs. To address this, Low-Rank Adaptation (LoRA)~\cite{LoRA} technology replaces full-parameter updates with two low-rank matrices, enabling efficient customization model training, thereby achieving hot-swappable style adaptation, allowing users to integrate new styles by downloading lightweight adapters.

\begin{figure}[t]
	\centering
	\includegraphics[width=0.475\textwidth]{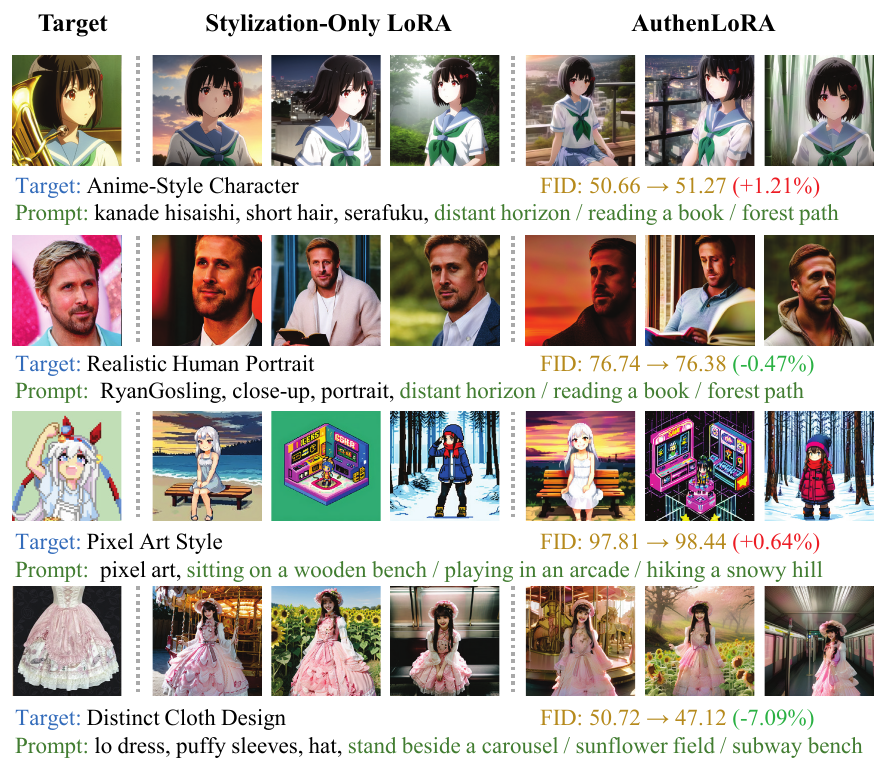}
	\caption{Visual results of stylization targets, images generated by stylization-only LoRA and watermarked images generated by AuthenLoRA. Our proposed AuthenLoRA preserves visual quality nearly indistinguishable from that of stylization-only LoRA-generated images.}
	\label{fig:Visual Results}
\end{figure}

However, the facilitation and widespread deployment of such stylized and personalized LoRA models also raise critical challenges regarding intellectual property (IP) protection and copyright enforcement~\cite{AICopyright}. LoRA models can be easily shared and redistributed in community and platform-based, often without proper attribution or authorization, which poses a significant risk of infringement and unauthorized commercial exploitation. Moreover, since LoRA is capable of generating highly stylized and personalized images, some malicious users may use the technology to forge identity information or create false news~\cite{FakePhotoTrump1,FakePhotoTrump2}, thus posing serious challenges to personal privacy and social security. 

\IEEEpubidadjcol

While watermarking technology could be implemented to protect the IP copyright of diffusion models and their generated images, most existing approaches~\cite{Tree-Ring, GaussianShading, AquaLoRA} mainly focus on base models. Recent works have proposed LoRA-specific watermarking method~\cite{BlackboxLoRA} based on backdoor watermarking, which verifies the copyright of the LoRA module by the specific designed input (i.e., trigger)-output behavior. However, these methods  focus solely on LoRA module copyright protection and fail to propagate watermarks to
the stylized images generated by LoRA module. Since traceability for LoRA-generated images is essential, there remains a critical gap in developing effective traceable watermarking mechanisms that are inherently integrated with the style-transfer generation process. 

Compared with base models, there are some challenges for watermarking  LoRA modules. Firstly, in contrast to base models that benefit from high-dimensional parameter spaces, LoRA's constrained trainable parameters present fundamental limitations for watermark embedding. Secondly, LoRA modules for diffusion modules are usually developed for specific tasks such as style transfer, while watermark embedding may disturb the learning of style representation, resulting undesirable artifacts in generated images. Furthermore, existing model watermarking techniques emphasize watermark extraction accuracy of watermarked models but suffer from high false positives, which misidentifies models without watermarks as watermarked ones, thereby mistakenly identifying them as pirated with a high possibility.

To address the research gap in LoRA traceable watermarking and overcome aforementioned challenges, we propose \textbf{AuthenLoRA}, a novel watermarking framework that embeds watermark into the parameters of the LoRA model, ensuring tight coupling with the stylized image generation process.  Our contributions can be summarized as follows:

\begin{itemize}
	\item We highlight the necessity of traceability watermarking for LoRA modules and propose \textbf{AuthenLoRA}, which simultaneously achieves stylized LoRA training and watermark embedding, ensuring that images generated by any models with the watermarked LoRA carry the watermark for traceability.
	\item We propose a dual-optimization framework of LoRA with expanded architecture and extended fine-tuning scope of ResNet blocks, which mitigates the inherent conflict between the learning of style distribution and the distribution shift augmented by the watermark, thereby concurrently achieving high-fidelity style transfer and robust watermark embedding.
	
	\item Different from previous approaches that only consider message regularization for watermarked images, we propose a zero-message regularization mechanism that enforces zero-sequence extraction from non-watermarked images, which effectively reduces false positive watermark detections compared with existing methods.
\end{itemize}

\section{Related Work}

\subsubsection{Text-to-Image Models}

Diffusion models~\cite{DDPM, ImprovedDDPM, DDIM} are a class of generative models~\cite{GAN, TRANSFORMER, VAE} that gradually transform Gaussian noise into complex data, such as images. Stable Diffusion~\cite{StableDiffusionRepository, StableDiffusionv2GithubRepository} is a latent diffusion model~\cite{LatentDiffusionPaper} that maps data to a compressed latent space, significantly improving computational efficiency. Owing to its impressive generative performance and open-source availability, Stable Diffusion has seen widespread adoption among the research community and digital artists, spurring a surge of derivative developments and custom adaptations.

\subsubsection{Low-Rank Adaptation (LoRA)}

To produce stylized or customized images, fine-tuning techniques such as Textual Inversion~\cite{TextualInversion}, ControlNet~\cite{ControlNet}, Dreambooth~\cite{Dreambooth}, and LoRA~\cite{LoRA} are employed. Among them, LoRA is widely used in the community~\cite{Civitai, HuggingFace} as a fine-tuning approach due to its favorable trade-off between adaptability and computational efficiency.

LoRA introduces low-rank trainable matrices to efficiently fine-tune large models, allowing them to adapt to downstream tasks while keeping most of the original model parameters frozen. In a standard neural network, consider a linear transformation \( y = \mathbf{W} x \), where \( \mathbf{W} \in \mathbb{R}^{d \times k} \) is a weight matrix. Fine-tuning the full matrix \( \mathbf{W} \) can be computationally expensive and parameter inefficient. LoRA approximates the update to \( \mathbf{W} \) as a low-rank decomposition:
\begin{equation}
	\label{Principle of LoRA}
	y = (\mathbf{W} + \alpha \cdot \Delta \mathbf{W})  x = (\mathbf{W} + \alpha \cdot \mathbf{A} \times \mathbf{B})  x
\end{equation}
$\text{where } \mathbf{A} \in \mathbb{R}^{d \times r}, \mathbf{B} \in \mathbb{R}^{r \times k},  r \ll \min(d, k).$
\( \alpha \) is a scaling factor to control the magnitude of the adaptation. During training, only \( \mathbf{A} \) and \( \mathbf{B} \) are updated, while \( \mathbf{W} \) remains frozen, significantly reducing the number of trainable parameters, making it possible to fine-tune large models on consumer hardware. The updated weight matrix can also be expressed as $\Delta \mathbf{W}$, which is the matrix product of two low-rank matrices $\mathbf{A} \in \mathbb{R}^{d \times r} \text{ and } \mathbf{B} \in \mathbb{R}^{r \times k}.$

As the core value of LoRA models lies in their ability to produce highly stylized outputs, it becomes imperative to develop copyright protection mechanisms that safeguard these generative identity models.

\subsubsection{Watermarking for Text-to-Image Models}
Traditional watermarking approaches~\cite{DWTDCT, DWTDCTSVD, RivaGAN, HiDDeN, TMM-Multilayer, TMM-DNNWatermark, TMM-FrPCET} embed traceable identifiers into generated images through post-processing operations, which lack integration with the generation pipeline and make them vulnerable to removal or evasion. 

To address these issues, approaches that embed watermark data during the image generation process are proposed. According to the stage at which the watermark is embedded, methods can be categorized as initial-noise-modification-based watermarking~\cite{Tree-Ring, GaussianShading, GaussianShading++, PRCWateramrk} and fine-tuning-based watermarking~\cite{StableSignature, VAEWMAdapter, CodecLatentWatermark, CodecRoSteALS, AquaLoRA, UNetmin2024watermark, TMM-PlugandUse}. However, these watermarking methods typically require intrusive modifications to the generation pipeline, which result in difficulty in migrating across different models or sampling pipelines.

Furthermore, existing watermarking methods for LoRA protect model or image copyright by embedding backdoors into the model~\cite{BlackboxLoRA, TMM-Backdoor} or fine-tuning the VAE decoder~\cite{VAELoRA} to embed watermark into the generated images. However, these methods suffer from loose coupling with the image generation process, and the embedded watermark information cannot be reconfigured after training, which limits their practical deployment on model distribution platforms requiring user traceability.

\section{Method}

\subsection{Problem formulation}
\begin{figure*}[t]
	\centering
	\includegraphics[width=\textwidth]{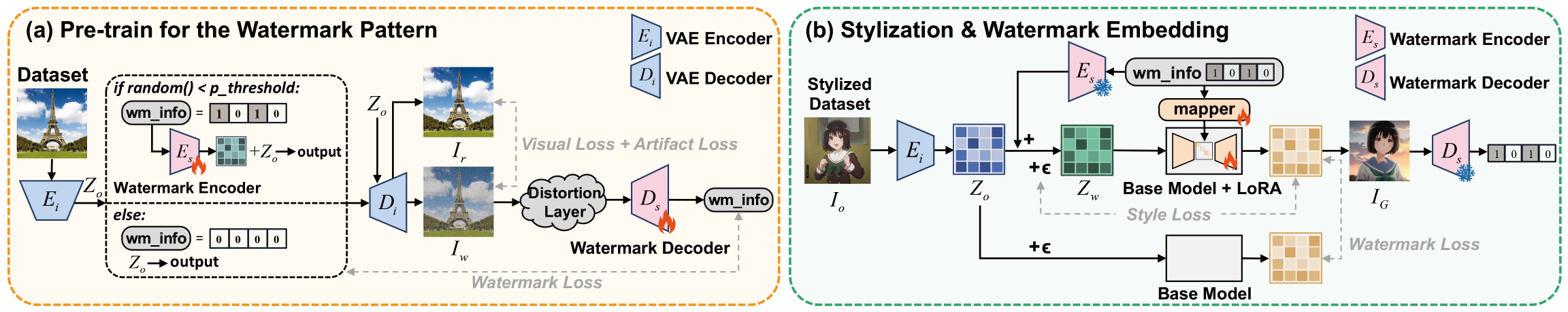}
	\caption{The overall framework of AuthenLoRA. Our proposed training pipeline consists of two stages. (a) The first stage is latent watermark paradigm pre-training. In this stage, we jointly train the watermark encoder $E_s$ and decoder $D_s$ in the latent space. (b) In the second stage, we jointly train LoRA and the watermark mapper. LoRA will learn both the pre-trained watermark paradigm and the stylized content. The watermark mapper is used to embed bit-level watermark message into LoRA.}
	\label{fig:Overall Method}
\end{figure*}

Our objective is to address the loose coupling between stylized image generation and watermark embedding in existing methods. To this end, we propose a unified framework that performs joint optimization of style transfer and watermark embedding. We formulate the problem as optimizing the LoRA module such that the noise predictor learns to capture both the watermark-induced bias and the stylization bias simultaneously. For stylization, we align the noise predicted by the U-Net merged with LoRA with the reference noise sampled from the training-time noise scheduler, by which LoRA can learn a distribution consistent with the domain of the training set. 
For watermarking, we expect the LoRA module to learn a watermark-specific pattern, which is required for reference. Therefore, we pretrain a watermarking framework to construct this pattern.
Thereby, as Fig.~\ref{fig:Overall Method} shows, the overall framework of the proposed method comprises a two-stage training pipeline. The first stage pretrains a watermark encoder-decoder for reference watermark pattern construction. The second stage is jointly training LoRA with watermarking objective and stylization objective.

\begin{figure}[t]
	\centering
	\includegraphics[width=0.45\textwidth]{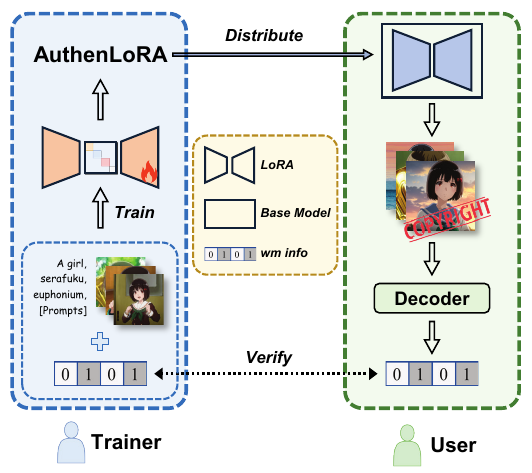}
	\caption{ The threat model. The LoRA trainer trains and owns the copyright of AuthenLoRA. Users use AuthenLoRA to generate stylized images with watermarks indicating their identity.}
	\label{fig:Threat Model}
\end{figure}

\subsection{Threat Model}
The threat model involves three entities: the LoRA trainer \textit{Alice}, the benign user \textit{Bob}, and the malicious user \textit{Eve}. The watermarking objective is to embed imperceptible identifiers into generated images while preserving visual quality. 

\begin{itemize}
	\item \textit{LoRA Trainer (Alice):} Alice constructs a dataset characterized by a unique artistic style and trains a LoRA model using our proposed AuthenLoRA. To  assert copyright ownership and facilitate provenance tracking, Alice embeds a unique watermark, serving as a specific user identifier, into the model prior to its distribution.
	\item \textit{Benign User (Bob):} Upon obtaining the usage license, Bob employs AuthenLoRA with the base model to generate stylized images $I_G$ from text prompts $\textit{c}$. The user identifier is tightly bound to AuthenLoRA. In other words, all stylized images generated using AuthenLoRA carry unique User-ID of the generator.
	\item \textit{Malicious User (Eve):} Acting as an adversary, Eve will use various image processing distortions and model-level modifications to remove any watermark information potentially concealed with both the generated images and the model. Eve's operation poses a challenge to the integrity and robustness of our proposed method.
\end{itemize}

\subsection{Pre-training for the Watermark Pattern}

In this training phase, we hope to train an image watermarking network that can generate latent watermark pattern. For this reason, we utilize the classical encode-decode network, which contains an encoder $E_s$ for watermark embedding and a decoder $D_s$ for watermark extraction. Following the design of AquaLoRA~\cite{AquaLoRA}, we choose to inject the watermark in the latent space rather than in the pixel space, as latent-space perturbations are easier to be learned by U-Net.
Our training objective comprises two primary goals: (1) to ensure consistency between the embedded and extracted watermark messages, and (2) to preserve the visual quality of the stylized outputs, avoiding perceptible degradation or watermark-induced artifacts. 

\subsubsection{Watermarking Loss} To ensure accurate watermark retrieval, Binary Cross Entropy (BCE) loss between the embedded watermark $m$ and its extracted counterpart $m'$ is usually implemented. However, current watermarking frameworks exhibit a critical asymmetry: they optimize for reliable watermark extraction while neglecting decoder behavior on clean inputs, often leading to false positives (decoding hallucinated messages from unwatermarked content).
To address this, we introduce a \textbf{zero-message regularization} mechanism, which incorporates clean images into training with a certain probability and assigns them an all-zero message as ground truth message, thereby enforcing a clear decision boundary between watermarked and non-watermarked images.
Specifically, during each training iteration, a Bernoulli sampling process with a predefined probability $p \in [0,1]$ determines whether a clean image is used and an all-zero message is assigned.

At each iteration, we first sample the target message $m$ as:
\begin{equation}
	m = 
	\begin{cases}
		\text{Random}\{0,1\}^k, & \text{with probability } 1 - p \\
		\mathbf{0}, & \text{with probability } p
	\end{cases}
	\label{eq:msg_sampling}
\end{equation}

Then, the corresponding latent input $z_w$ is defined by:
\begin{equation}
	z_w = 
	\begin{cases}
		z_o+E_{s}(m), & \text{if } m \ne \mathbf{0} \\
		z_o, & \text{if } m = \mathbf{0}
	\end{cases}
	\label{eq:latent_construction}
\end{equation}

The encoder and decoder is then trained using the standard BCE loss between the predicted message $\hat{m}$ and the assigned target $m$, yielding the following loss formulation:
\begin{equation}
	\mathcal{L}_{msg} = \mathbb{E}_{z,m}[\text{BCE}(\hat{m},m)]
\end{equation}

\begin{figure}[t]
	\centering
	\includegraphics[width=0.475\textwidth]{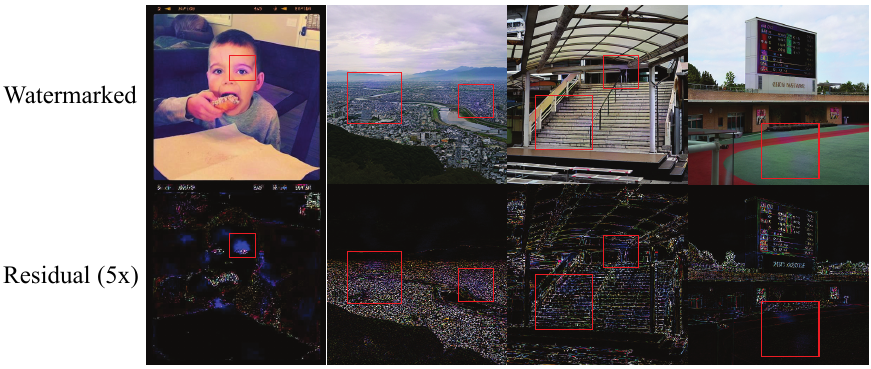}
	\caption{The watermarked images and their corresponding residuals generated by watermark encoder trained without the PRTS loss. Regions exhibiting high-saturation artifacts are highlighted. The residual intensity has been amplified 5x to enhance visibility.}
	\label{fig:AquaLoRA_Residual}
\end{figure}

\subsubsection {Visual Fidelity Loss} To preserve visual fidelity post-embedding, we utilize the LPIPS loss~\cite{LPIPS} and Peak Regional Variation Loss (PRVL)~\cite{AquaLoRA} between the watermarked image $I_w$ and its VAE-reconstructed version $I_r$. {LPIPS} computes the perceptual similarity and aligns more closely with human visual perception. {PRVL} computes the most salient regions of local variation within the image to constrain the maximum strength of the watermark signal. Furthermore, as shown in Fig.\ref{fig:AquaLoRA_Residual}, we observe that the aforementioned loss functions fail to suppress the high-saturation watermark residuals introduced by the watermark encoder, which compromises the invisibility of the watermark and degrades the visual quality of the generated images. To address this, we propose the \textbf{Perceptual Residual Tone Suppression (PRTS)} loss, which selectively suppresses chromatic anomalies in perceptually sensitive hue ranges.
It first converts the residual image $R$ to HSV space differentiably to explicitly separate hue and saturation components, then applies a soft binary mask to identify target hue bands. The loss aggregates saturation values within these bands to penalize residual energy:

\begin{equation}
	\mathcal{L}_{PRTS} = \frac{1}{N}\sum _{i=1}^{N} M_i \cdot S_i
\end{equation}
where
\begin{itemize}
	\item $R=\text{clip}(\frac{I^w-I^c+1}{2},0,1)$ is the normalized residual image,
	\item $(H_i, S_i, V_i)=\text{HSV}(R_i)$ denotes the differentiable HSV representation of the residual pixel $i$,
	\item $M_i = 1_{[h_{\text{low}},h_{\text{high}}]}(H_i)$ is a soft mast with H-channel range of watermark residual
	\item $S_i$ is the saturation component of HSV representation,
	\item $N$ is the total number of pixels.
\end{itemize}

This formulation encourages the model to avoid generating high-saturation residuals within the visually sensitive hue range, thereby enhancing imperceptibility without overly penalizing the entire residual spectrum.

In summary, the overall loss function can be summarized as:
\begin{equation}
	\mathcal{L}_{Total} = \mathcal{L}_{msg} + \lambda \mathcal{L}_{LPIPS} + \mu \mathcal{L}_{PRVL} + \nu \mathcal{L}_{PRTS}
\end{equation}
where $\lambda, \mu \text{ and } \nu$ are hyperparameters, which will be discussed in detail in the supplementary matarial.
The well-trained $E_s$ will serve to embed the watermark information into the latent space of clean images, which will later be used to guide the training of LoRA. The decoder $D_s$ will be implemented for watermark extraction from generated images. Detailed training pipelines and corresponding pseudocode are provided in the supplementary material.

\subsection{Stylization and Watermark Embedding}

\subsubsection{LoRA Parameter Modulation via Diagonal Scaling}

In contrast to conventional LoRA, which introduces low-rank updates $ \Delta \mathbf{W} = \mathbf{A} \times \mathbf{B}$ to a frozen base model, we extend this structure by injecting a diagonal scaling matrix $\mathbf{S}$ into the LoRA formulation:
\begin{equation}
	\label{scaling matrix}
	\Delta \mathbf{W} = \mathbf{A} \times \mathbf{S} \times \mathbf{B}
\end{equation}
where $\mathbf{S} \in \mathbb{R} ^ {r \times r}$ is a learnable matrix determined by an auxiliary mapper network. This design decouples watermark-specific information from the low-rank components $\mathbf{A}$ and $\mathbf{B}$, allowing LoRA to learn a stable style-transfer prior while enabling dynamic watermark injection via matrix $\mathbf{S}$.
Given a binary secret message vector $m \in \left\{0,1 \right\}^k$, a lightweight neural mapper $f_\phi$ encodes this vector into a continuous embedding $\mathbf{v} \in \mathbb{R}^r$, which is then lifted to a diagonal matrix:
\begin{equation}
	\mathbf{S} = \text{diag}(f_\phi(m)).
\end{equation}

The scaling matrix is then applied to modulate the LoRA down-projection weights at each layer. We choose the down-projection layers of LoRA because these layers play a more prominent feature extraction role for the final generated image and have less impact on the final image quality. Importantly, since $\mathbf{S}$ is generated at runtime, this formulation enables on-demand watermark customization, allowing a well-trained LoRA to support diverse watermark payloads without retraining.

\subsubsection{LoRA Fine-tuning Scope}
Conventional LoRA fine-tuning typically targets the Transformer blocks, as they are more sensitive to global features and semantic understanding essential for stylization tasks, thereby achieving a balance between training efficiency and stylization performance. However, we observe that this fine-tuning scope leads to slow convergence in watermark embedding tasks, since the fine-tuned components lack sufficient attention to rich texture patterns critical for watermark representation. 

To address this limitation, we extend the fine-tuning scope to include ResNet blocks, leveraging their enhanced sensitivity to structural and textural details to improve both training efficiency and watermark effectiveness.  Since the pre-trained watermark encoder embeds the watermark within the latent space, the ResNet modules are positioned to directly capture the intricate details of these latent watermark patterns. The controlled increase in model size and significantly accelerating training convergence make this design particularly well-suited for practical deployment and community sharing. We further validate this improved design in experiments.

\subsubsection{Watermark Embedding}

At this stage, our goal is to fine-tune the U-Net via LoRA to internalize the watermark pattern obtained during the pre-training phase. By embedding watermarks into the training dataset to alter its data distribution and subsequently fine-tuning the U-Net on this modified dataset, the U-Net is encouraged to learn the watermark pattern embedded within the data distribution. 

Specifically, a watermark perturbation, which is determined solely by the embedded message $m$, is artificially introduced into the clean latent representation $\mathbf{Z}_o$, resulting in a watermarked latent representation $\mathbf{Z}_w$ via the pretrained watermark encoder $E_s$. To embed this perturbation into the LoRA weights, the model must learn the correspondence between the watermark-induced distortion and the deviation in noise prediction. During interference, the LoRA module is expected to reproduce this learned deviation, thereby injecting the pretrained watermark pattern into the output image. The pre-trained watermark decoder $D_s$ is employed to extract the embedded watermark from generated images.

To achieve this, we obtain both the original noise prediction from the base model and the watermark-shifted noise prediction at each training step. 
For the watermark-shifted noise prediction, the watermarked latent $\mathbf{Z}_w$ is perturbed by additive Gaussian noise $\boldsymbol{\epsilon} \sim \mathcal{N}(0, \sigma^{2} I)$, as introduced by the noise scheduler. Subsequently, the noisy latent is passed through the LoRA-merged diffusion model $\mathcal{D}_{\theta+\Delta W}$ with timestep $t$ and text condition $c$ to predict the noise, which can be expressed as:

\begin{equation}
	\mathbf{Z}_{\text{pred\_wm}} = \mathcal{D}_{\theta+\Delta W} (\sqrt{\bar{\alpha}_t} \mathbf{Z}_w + \sqrt{1 - \bar{\alpha}_t} \boldsymbol{\epsilon}, t, c)
\end{equation}

For the original noise prediction, the original latent $\mathbf{Z}_o$ is perturbed with the same noise $\boldsymbol{\epsilon}$, and the processed by the frozen base model $\mathcal{D}_\theta$ without LoRA:
\begin{equation}
	\mathbf{Z}_{\text{pred\_clean}} = \mathcal{D}_\theta (\sqrt{\bar{\alpha}_t} \mathbf{Z}_0 + \sqrt{1 - \bar{\alpha}_t} \boldsymbol{\epsilon}, t, c)
\end{equation}

We define the watermark loss as the mean squared error (MSE) between the watermark-shifted noise prediction and the original noise prediction:
\begin{equation}
	\mathcal{L}_{wm} = \left \| \mathbf{Z}_{\text{pred\_wm}} - \mathbf{Z}_{\text{pred\_clean}} \right \|^2,
\end{equation}
which explicitly guides the LoRA module to encode the watermark signal through its modulation of the noise prediction process.

\subsubsection{Style Migration}
To ensure that the LoRA model effectively captures the distribution of the stylized dataset, it is essential to align the predicted noise from the merged model, which consists of the training LoRA model  $\Delta W$ and the frozen base model $\theta$, with the reference noise generated by noise scheduler during training. To this end, we compute the mean squared error (MSE) between the generated reference noise $\boldsymbol{\epsilon}$ and the predicted noise $\mathbf{Z}_{\text{pred\_wm}}$, defined as:
\begin{equation}
	\mathcal{L}_{style} = \left \| \mathbf{Z}_{\text{pred\_wm}} - \boldsymbol{\epsilon} \right \|^2,
\end{equation}
which guides and constrains the LoRA module toward learning particular stylization consistent with the target dataset distribution.

\subsubsection{Training with Multi-Objective Loss}In summary, the total loss is a weighted sum of the above terms:
\begin{equation}
	\mathcal{L}_{total} = \lambda_{style} \mathcal{L}_{style} + \lambda_{wm} \mathcal{L}_{wm},
\end{equation}
with $\lambda_{style} \text{ and } \lambda_{watermark}$ are hyperparameters. 
In training, we adopt a dynamic weighting strategy that adjusts $\lambda_{style}$ and $\lambda_{wm}$ over the course of training. 
In the early training phase, we emphasize watermark fidelity by setting $\lambda_{wm} \gg \lambda_{style}$. Once the watermark bitwise decoding accuracy surpasses a predefined threshold, we gradually increase $\lambda_{style}$ to enhance the stylization quality, thereby ensuring a balanced trade-off between visual expressivity and watermark decodability.

\section{Experiments}

\begin{table*}[t]
	\centering
	\caption{Comparison of the performance of our proposed method with other watermarking methods based on post-processing and during the diffusion process. }
	\begin{tabular}{llcccccccc}
		\toprule
		\multicolumn{1}{c}{\multirow{2}{*}{Method}} & \multicolumn{1}{c}{\multirow{2}{*}{Capacity}} & \multicolumn{4}{c}{Fidelity}                                                              & \multirow{2}{*}{BitAcc} & \multirow{2}{*}{TPR} & \multirow{2}{*}{BitAcc(Adv.)} & \multirow{2}{*}{TPR(Adv.)} \\ \cmidrule{3-6}
		\multicolumn{1}{c}{}                        & \multicolumn{1}{c}{}                          & CLIP$\uparrow$       & FID$\downarrow$      & NIQE$\downarrow$     & PIQE$\downarrow$     &                         &                      &                               &                            \\ \midrule
		Original Dataset                            & --                                            & 33.2190              & 68.9825              & 3.9185               & 42.5618              & --                      & --                   & --                            & --                         \\ \midrule
		DWT-DCT                                     & 48 bits                                       & 33.1381              & 70.0182              & 3.8879               & 42.3636              & 0.8553                  & 0.6710               & 0.5536                        & 0.1295                     \\
		DWT-DCT-SVD                                 & 48 bits                                       & 33.3341              & 70.3459              & 3.9080               & 42.4197              & 0.9794                  & 0.9625               & 0.6469                        & 0.2599                     \\
		RivaGAN                                     & 32 bits                                       & 33.0010              & 72.0922              & 3.7961               & 41.2612              & \textbf{0.9927}         & 0.9935               & 0.8322                        & 0.6238                     \\ \midrule
		Tree-Ring                                   & 1 bit                                         & \textbf{34.3225}     & 69.8833              & 4.2457               & 45.0475              & -                       & 0.9440               & -                             & 0.6209                     \\
		Stable Signature                            & 48 bits                                       & 33.1384              & 71.4579              & \textbf{3.6718}      & 42.4512              & 0.9914                  & \textbf{1}           & 0.7838                        & 0.5722                     \\
		AquaLoRA                                    & 48 bits                                       & 32.9027              & 68.3604              & 3.9358               & 45.4872              & 0.9359                  & 0.9140               & 0.8938                        & 0.8061                     \\
		\textbf{Ours (AuthenLoRA)}                       & 48 bits                                       & 33.7285              & \textbf{68.3025}     & 3.7782               & \textbf{35.4764}     & 0.9885                  & \textbf{1}           & \textbf{0.9525}               & \textbf{0.9622}            \\ \bottomrule
	\end{tabular}
	\label{tabel:Overall Comparison in Fidelity and Accuracy}
\end{table*}

\subsection{Implementation Details}

To demonstrate the generalizability and expressiveness of our approach across different stylization domains, we train four distinct LoRA models, each corresponding to a unique stylization type: (1) an anime-style character, (2) a realistic human portrait, (3) a specific artistic style, and (4) a distinct clothing design. All LoRA modules are trained based on the Stable Diffusion 1.5 base model. Visual demonstrations, training configurations and baseline implementation details are provided in the supplementary material.

\begin{table*}[]
	\centering	
	\caption{The results of combining different with base models of different stylistic tendencies to generate images and evaluate bit accuracy and TPR@FPR=$10^{-6}$.}
	\begin{tabular}{lcccccc} 
		\toprule
		\multicolumn{1}{c}{\multirow{2}{*}{LoRA Model Style}}           & \multicolumn{6}{c}{Base Model Style} \\
		\cmidrule{2-7}
		\multicolumn{1}{c}{}                                            & Realistic 1                                                               & Realistic 2                                                              & Realistic 3                                                               & Cartoonish 1                   & Cartoonish 2                   & Cartoonish 3                    \\ 
		\midrule
		Anime Character (Cartoonish)                                                 & 0.9902 / \textbf{1}                                              & {\cellcolor[rgb]{0.937,0.937,0.937}}\textbf{0.9933 /~\textbf{1}} & 0.9842 / \textbf{1}                                              & 0.9916 / \textbf{1}  & 0.9909 / \textbf{1}  & 0.9778 / \textbf{1}   \\
		Realistic Character (Realistic)                                             & {\cellcolor[rgb]{0.937,0.937,0.937}}\textbf{0.9910 /~\textbf{1}} & 0.9764 /\textbf{ 1}                                              & 0.9750 / 0.998                                                   & 0.9612 / 0.998       & 0.9659 / 0.998       & 0.9612 / 0.998        \\
		Pixel Art Style (Cartoonish)                                                & 0.9817 / \textbf{1}                                              & {\cellcolor[rgb]{0.937,0.937,0.937}}0.9842 / \textbf{\textbf{1}} & 0.9751 / 0.998                                                   & \textbf{0.9870 / 1}  & 0.9587 / 0.996       & 0.9628 / 0.994        \\
		Clothing Design (Realistic)                                                 & 0.9605 / 0.998                                                   & 0.9460 / 0.996                                                   & {\cellcolor[rgb]{0.937,0.937,0.937}}\textbf{0.9838 /~\textbf{1}} & 0.9324 / 0.982       & 0.9511 / 0.998       & 0.9615 / \textbf{1}   \\
		\bottomrule
	\end{tabular}
	\label{table:LoRA with different base model}
\end{table*}

\begin{figure*}
	\centering
	\includegraphics[width=\textwidth]{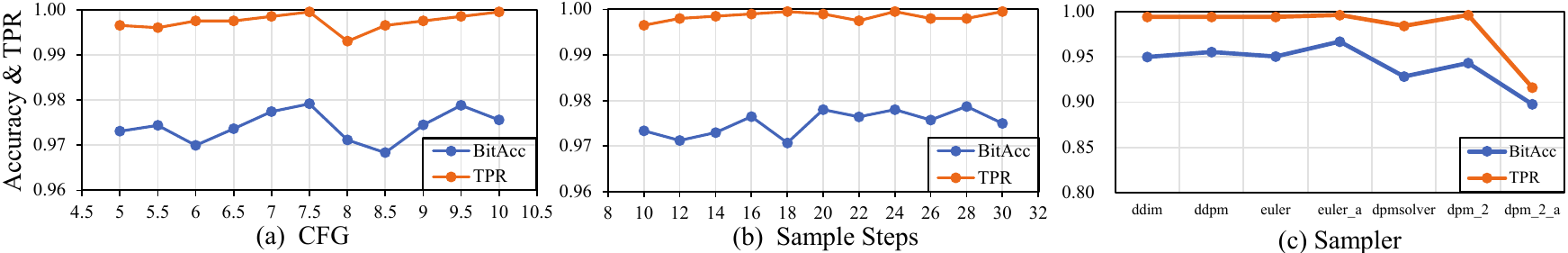}
	\caption{Effect of generation parameters ((a) Classifier Free Guidance, (b) Sample Steps and (c) Sampler) on bit accuracy and TPR@FPR=$10^{-6}$ during image generation for Stable Diffusion.}
	\label{fig:Generation_Parameters}
\end{figure*}

\subsection{Fidelity, Accuracy and Robustness of Generated Images}

Figure~\ref{fig:Visual Results} presents a visual overview of the target styles, the images generated by stylization-only LoRA and our proposed AuthenLoRA. As shown, our method effectively preserves stylization fidelity—such as facial characteristics, pixelart artistic styles, and distinctive clothing designs—thus maintaining visual coherence and aesthetic quality.

Table~\ref{tabel:Overall Comparison in Fidelity and Accuracy} presents a comprehensive comparison of seven watermarking methods in terms of visual fidelity, watermark extraction accuracy, and robustness under adversarial conditions. We control the FPR at $10^{-6}$ and calculate the TPR. Since Tree-Ring is a zero-bit watermark method, the bit accuracy cannot be calculated, and we only calculated the TPR. The best results are presented in bold text. For robustness evaluation, we apply common image processing operations (e.g., JPEG compression, Gaussian blur, etc.) on the watermarked images obtained by different methods and evaluate their performance.

Post-processing-based methods generally achieve relatively low FID and high CLIP scores, indicating minimal degradation in visual quality. Moreover, they demonstrate high bit accuracy (BitAcc) and true positive rate (TPR) values under benign conditions. However, these methods exhibit a pronounced drop in robustness when subjected to common image processing attacks. Methods based on the during-diffusion process perform better in terms of robustness, but cause large degradation in image quality.

Our method, AuthenLoRA, achieves the best overall performance. It preserves nearly lossless visual quality compared to the original dataset in non-adversarial settings, while attaining near-perfect watermark extraction accuracy. Under adversarial conditions, it significantly outperforms all baseline methods in both bit accuracy and true positive rate, demonstrating superior robustness of our proposed method. Detailed per-category results for each of the four stylized LoRA models are provided in the supplementary material.

\subsection{Universality}
\subsubsection{AuthenLoRA with Different Base Model}

To further investigate the generalization capability of the proposed AuthenLoRA models across different base models, we conducted cross-combination experiments between LoRA models trained on diverse style datasets and base models with varying stylistic tendencies. In this experiment, both the LoRA models and base models are categorized into two major stylistic groups: \textit{realistic} and \textit{cartoonish}. Table~\ref{table:LoRA with different base model} summarizes the training–testing pairings and experimental results. The gray underlining in the table indicates that this AuthenLoRA was trained under this base model. The best results are highlighted in bold.

As can be observed, the watermarking performance generally reaches its optimum when the AuthenLoRA is paired with the base model it was trained on. Nevertheless, even in cross-style combinations, the AuthenLoRA models are able to maintain reasonably high watermarking accuracy and true positive rates, which proves that users have the flexibility to choose between different base models based on creativity or demand without substantially affecting the robustness or effectiveness of the embedded watermarks.

\subsubsection{Generation Parameters}
To evaluate the generalization capability of AuthenLoRA, we examine the effectiveness of watermark on the influence of image generation parameters—including the classifier-free guidance scale (CFG), the number of sampling steps, and the choice of sampler.

As shown in Fig.~\ref{fig:Generation_Parameters}~(a) and~(b), varying the CFG scale and sampling steps over a wide range has negligible impact on the watermarking performance. Across all configurations, the bit accuracy consistently remains above 97\%, while the TPR remains stably above 99\%, indicating strong resilience of the watermark to typical generation parameter fluctuations. Furthermore, Fig.~\ref{fig:Generation_Parameters} (c) illustrates the watermark performance across different samplers. Despite the inherent variation in the sampling dynamics introduced by different algorithms, the BitAcc stays above 96\% and the TPR similarly maintains a high level across all samplers.

\begin{figure*}
	\centering
	\includegraphics[width=\textwidth]{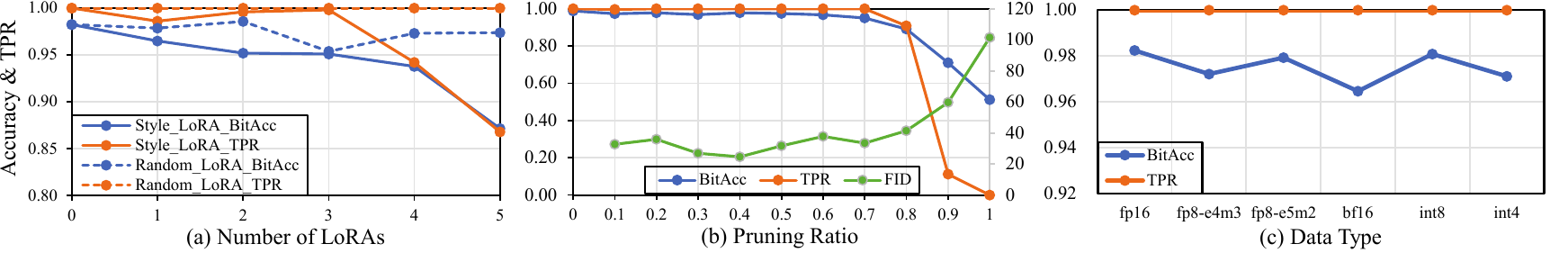}
	\caption{Robustness evaluation of AuthenLoRA against three model-level modifications: (a) LoRA merging, (b) model pruning and (c) model quantization.}
	\label{fig:Robustness}
\end{figure*}
	
\subsubsection{LoRA Variants}

To further validate the effectiveness and robustness of the proposed watermarking method, we compare the watermark performance between two training paradigms: (1) separately training a stylization LoRA and a watermarking LoRA, and subsequently applying them jointly during image generation; and (2) our AuthenLoRA framework, which performs joint training of stylization and watermarking objectives. In addition, we examined the performance of the watermarking approach across different LoRA variants, including \textit{LoCon}, \textit{LoHa}, and \textit{LoKr}~\cite{LyCORIS, LyCORISGithub}, to assess its generalizability.

The experimental results are summarized in Table~\ref{table:LoRA Variants}. The best results are highlighted in bold. It can be observed that the AuthenLoRA framework consistently achieves superior watermarking performance across all LoRA variants. Even on more challenging variants such as LoHa and LoKr, AuthenLoRA maintains a robust performance, while the performance of individually trained watermarking LoRAs on the same configurations degrades. This demonstrates the strong generalizability and compatibility of our proposed method, and that the joint training strategy facilitates a better integration of stylization and watermarking objectives.

\begin{table}[]
	\centering
	\caption{Comparison of different variants of stylized LoRA and watermarked LoRA.}
	\begin{tabular}{lllll} 
		\toprule
		\multicolumn{1}{c}{\multirow{2}{*}{Style LoRA}} & \multicolumn{4}{c}{Watermark LoRA}                                                                          \\ 
		\cmidrule{2-5}
		\multicolumn{1}{c}{}                            & \multicolumn{1}{c}{LoRA} & \multicolumn{1}{c}{LoCon} & \multicolumn{1}{c}{LoHa} & \multicolumn{1}{c}{LoKr}  \\ 
		\midrule
		None                                            & 0.9585 / \textbf{1}      & 0.9695 / \textbf{1}       & 0.9037 / 0.89            & 0.8891 / 0.92             \\
		LoRA                                            & 0.9611 / \textbf{1}      & 0.9665 / \textbf{1}       & 0.8720 / 0.77            & 0.8890 / 0.92             \\
		LoCon                                           & 0.9723 / \textbf{1}      & 0.9733 / \textbf{1}       & 0.9167 / 0.93            & 0.8974 / 0.97             \\
		LoHa                                            & 0.9661 / \textbf{1}      & 0.9730 / \textbf{1}       & 0.8050 / 0.45            & 0.8763 / 0.90             \\
		LoKr                                            & 0.9664 / \textbf{1}      & 0.9684 / \textbf{1}       & 0.8705 / 0.78            & 0.8958 / 0.96             \\
		AuthenLoRA                                      & \textbf{0.9885 / 1}      & \textbf{0.9948 / 1}       & \textbf{0.9266 / 0.99}   & \textbf{0.9300 / 0.98}    \\
		\bottomrule
	\end{tabular}
	\label{table:LoRA Variants}
\end{table}
	
\subsection{Robustness of AuthenLoRA}
In real-world scenarios, potential attackers may deploy model-level attacks against AuthenLoRA to remove the embedded watermark. To assess the robustness of AuthenLoRA under these conditions, we analyze its performance in three scenarios: LoRA merging, model pruning and model quantization. The watermark's efficacy is evaluated based on bit accuracy and TPR@FPR=$10^{-6}$. The experimental results are presented in Fig.~\ref{fig:Robustness}.

\subsubsection{LoRA Merging}
In practical applications of stylized image generation, digital artists frequently merge multiple LoRA models to achieve a superimposed effect that combines various stylistic tendencies. This merging process, however, poses a potential threat, as it could overwrite or corrupt the watermark information embedded by AuthenLoRA.

To simulate this potential attack, we first evaluated the performance of watermark after merging AuthenLoRA with multiple publicly available stylized LoRA models sourced from model platform Civitai~\cite{Civitai}. Additionally, inspired by LoRAGuard~\cite{BlackboxLoRA}, we conducted a second assessment by merging AuthenLoRA with randomly initialized LoRA models.

As illustrated in Fig.~\ref{fig:Robustness}(a), AuthenLoRA maintains bit accuracies exceeding 90$\%$ and 97$\%$, respectively, even after being merged with five stylized LoRA models and five randomly initialized LoRA models. These result demonstrates that AuthenLoRA retains significant robustness, even in complex application scenarios involving model composition.

\subsubsection{Model Pruning}
An attacker may attempt to disrupt the embedded watermark by pruning (i.e., zeroing out) model parameters, while simultaneously aiming to preserve the LoRA's core stylization functionality. To simulate this attack vector, we apply magnitude pruning to the AuthenLoRA model. Given a pruning ratio $p \in [0, 1]$, the algorithm ranks all parameters by their absolute magnitude and eliminates the smallest $p$ proportion of them by setting them to zero.

As illustrated in Fig.~\ref{fig:Robustness}(b), AuthenLoRA maintains a bit accuracy of $95\%$ and a $100\%$ TPR even at a pruning ratio of $p=0.7$. However, as $p$ increases further, the FID score rises sharply, indicating that higher pruning ratios simultaneously destroy the core stylization functionality while attempting to erase the watermark. This result demonstrates that an attacker cannot leverage this method to remove the watermark without also compromising the model's intended stylization.

\subsubsection{Model Quantization}
Model quantization, a common technique for compressing models, represents another potential attack vector. The process involves converting model parameters to lower-precision data types (e.g., 8-bit or 4-bit integers), which could inadvertently alter or erase the embedded watermark information. To investigate this threat, we simulated the attack by quantizing the AuthenLoRA model into various data formats, including fp8, bf16, int8, and the highly aggressive int4. 

As shown in Fig.~\ref{fig:Robustness}(c), the watermark exhibits exceptional resilience. Across all tested quantization levels, AuthenLoRA consistently achieves near-perfect bit accuracy and a TPR of $100\%$. This result demonstrates that the watermark is robust to precision reduction and remains fully detectable even after aggressive model compression.

\subsection{Ablation Study}

\begin{table}[]
	\centering	
	\caption{Performance comparison of six watermark decoders on a binary classification task, using 500 watermarked and 500 non-watermarked images.}
	\begin{tabular}{lccc}
		\toprule
		\multicolumn{1}{c}{Method} & Precision            & Recall               & FPR                  \\ \midrule
		DWT-DCT                    & 0.6218               & 0.8960               & 0.5690               \\
		DWT-DCT-SVD                & 0.5490               & 0.9810               & 0.8365               \\
		RivaGAN                    & 0.5000               & \textbf{1.0000}      & 1.0000               \\ 
		Stable Signature           & 0.5000               & \textbf{1.0000}      & 1.0000               \\
		AquaLoRA                   & 0.5000               & \textbf{1.0000}      & 1.0000               \\ \midrule
		w/o Zero-Regularization    & 0.5000               & \textbf{1.0000}      & 1.0000               \\
		w/ Zero-Regularization     & \textbf{0.9887}      & \textbf{1.0000}      & \textbf{0.0115}      \\ \bottomrule
	\end{tabular}
	\label{table:Zero-Regularization}
\end{table}

\subsubsection{Zero-Message Regularization}
To simulate a realistic environment and quantitatively evaluate the reliability of different watermark decoders, we conduct a binary classification experiment. For each method, we measure Precision, Recall and FPR, based on whether the decoder correctly identifies the presence or absence of a watermark. An image is considered non-watermarked if the decoded message is an all-zero or all-one vector.

As shown in Table~\ref{table:Zero-Regularization}, all baseline methods, including both post-process and during-diffusion methods, achieve relatively high recall, indicating that they reliably detect watermarks when present. But most of them suffer from poor precision and high FPR, indicating that they frequently produce hallucinated watermarks on clean images. 

In contrast, our method with zero-message regularization achieves perfect recall, showing that the model is able to successfully recognize all images with watermarks. Meanwhile, our method significnatly improves the precision and reduces the FPR, which clearly demonstrates the benefit in suppressing spurious watermark decoding on non-watermarked content. Without zero-message regularization mechanism, the decoder falls back to the baseline behavior of always predicting a watermark, resulting in poor FPR and precision. Detailed results are discussed in the supplementary material.

\subsubsection{Watermark Artifact}

To quantitatively assess the impact of watermark embedding on image visual quality, we compare six watermark methods in terms of Peak Signal-to-Noise Ratio (PSNR) and Structural Similarity Index (SSIM). 

\begin{table}[]
	\centering
	\caption{Comparison of the strength of watermark residuals across six watermarking methods. }
	\begin{tabular}{lcc}
		\toprule
		\multicolumn{1}{c}{Method} & PSNR↑                & SSIM↑                \\ \midrule
		DWT-DCT                    & 38.6900              & 0.9687               \\
		DWT-DCT-SVD                & 38.4675              & \textbf{0.9762}      \\
		RivaGAN                    & \textbf{40.6875}     & \textbf{0.9762}      \\
		Tree-Ring                  & 12.2100              & 0.4822               \\
		Stable Signature           & 29.2475              & 0.8819               \\
		AquaLoRA            & 29.0925              & 0.9011               \\ \midrule
		w/o PRTS Loss              & 29.3550              & 0.9028               \\
		w/ PRTS Loss               & \textbf{29.6925}     & \textbf{0.9049}     		\\ \bottomrule
	\end{tabular}
	\label{table:Watermark Artifact}
\end{table}

\begin{figure}
	\centering
	\includegraphics[width=0.475\textwidth]{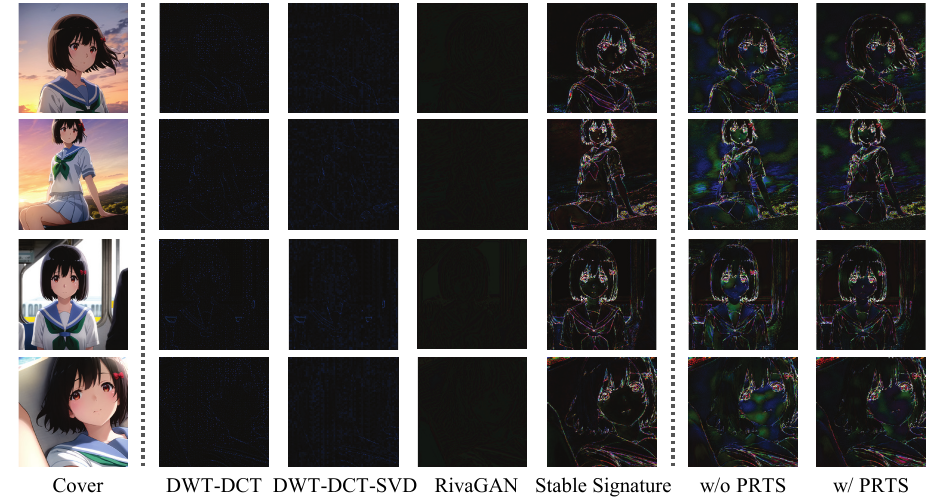}
	\caption{Visual Demonstration of watermark residuals from different methods. The residual intensity has been amplified 10x to enhance visibility.}
	\label{fig:Residual}
\end{figure}

Table~\ref{table:Watermark Artifact} presents quantitative result and Fig~\ref{fig:Residual} provides visual demonstration. For AquaLoRA and AuthenLoRA, we evaluate them using pre-trained watermark encoder. The best results for each of the two types of watermarking methods are highlighted in bold. It can be observed that post-processing watermarking methods generally achieve superior visual quality compared to during-diffusion approaches. In contrast, AquaLoRA exhibits visible watermark artifacts in generated images, which is reflected in its relatively lower PSNR and SSIM. Our method significantly mitigates such visual distortions and performs best among the watermarking methods based on during-diffusion process. Specifically, by introducing our proposed PRTS loss during training, we observe a clear improvement in visual quality compared to the variant without PRTS loss. This demonstrates that PRTS loss effectively reduces watermark residuals, thereby enhancing the invisibility of embedded watermark while maintaining desirable image fidelity.

\subsubsection{LoRA Layer Selection}

To investigate how different LoRA fine-tuning scopes influence the learnability and performance of embedded watermarks, we evaluate the performance of watermark by varying the subset of layers within the U-Net backbone subject to LoRA adaptation. 

\begin{figure}
	\centering
	\includegraphics[width=0.43\textwidth]{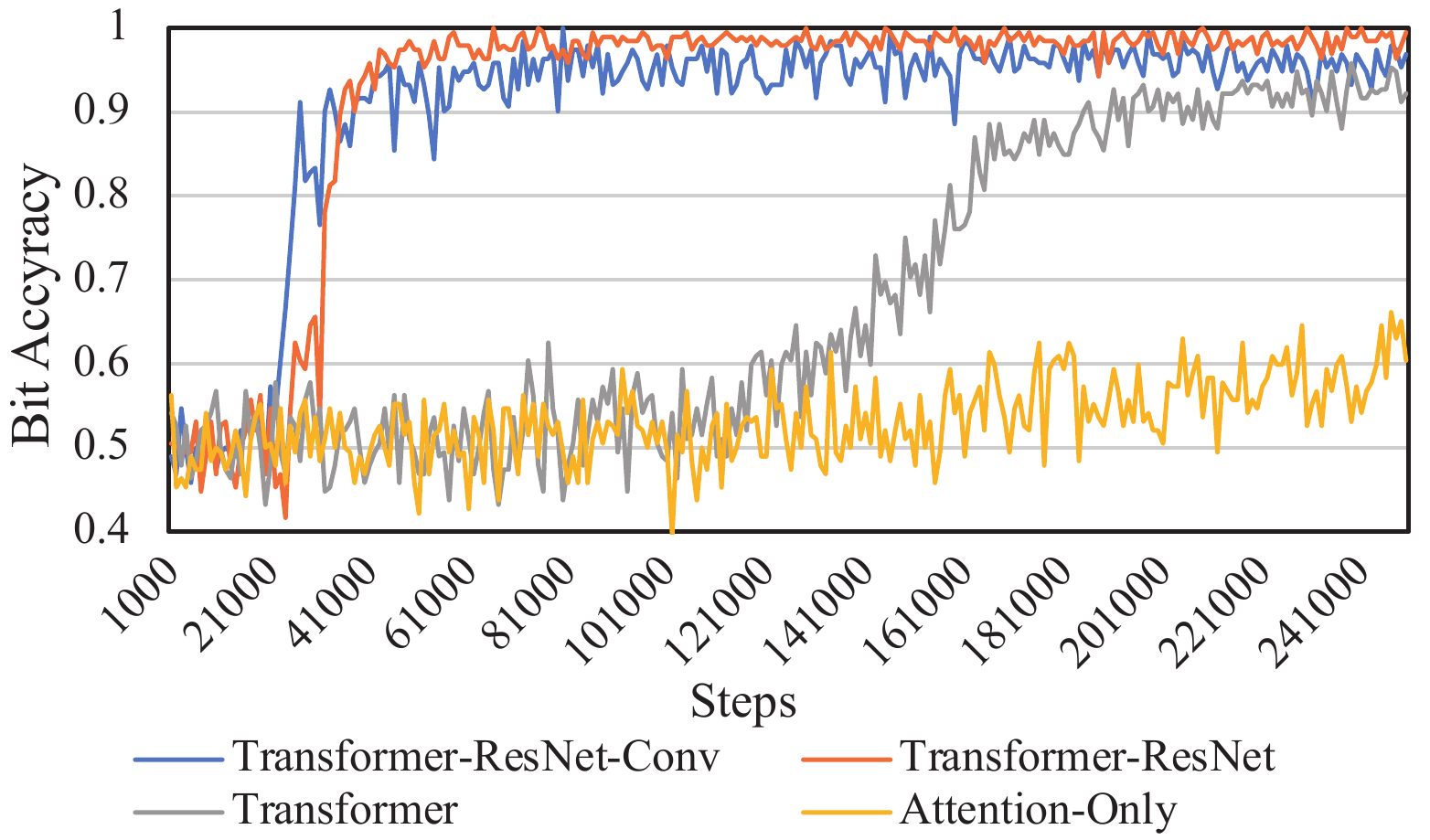}
	\caption{Watermark bit accuracy over training steps for different fine-tuning scopes on the U-Net backbone.}
	\label{fig:LoRA Layer Selection}
\end{figure}

As shown in Figure~\ref{fig:LoRA Layer Selection}, models with broader layer coverage—particularly those including ResNet blocks—exhibit significantly faster convergence in terms of bit-level watermark accuracy during training. We infer that the stacked residual blocks can accumulate multi-scale feature, making networks particularly sensitive to subtle signal variations and accelerating convergence of watermark learning. 

\begin{table}[]
	\centering
	\caption{Quantitative comparison of different LoRA fine-tuning scopes.}
	\begin{tabular}{llll}
		\toprule
		\multicolumn{1}{c}{Layer of LoRA} & \multicolumn{1}{c}{BitAcc} & \multicolumn{1}{c}{TPR} & \multicolumn{1}{c}{File Size (fp16)} \\ \midrule
		Transformer-ResNet-Conv                & 0.9792                     & 1                       & 747.28 MB                       \\
		Transformer-ResNet                     & 0.9874                     & 1                       & 413.25 MB                     \\
		Transformer                            & 0.9369                     & 0.9969                  & 360.11 MB                     \\
		Attention-Only                         & 0.5695                     & 0                       & 166.72 MB                     \\ \bottomrule
	\end{tabular}
	\label{table:LoRA Layer Selection}
\end{table}

Table~\ref{table:LoRA Layer Selection} quantifies the bit accuracy, TPR and model size. While \textit{Transformer-ResNet-Conv} fine-tuning achieves the highest BitAcc and perfect detection, it also incurs the largest model file size. Excluding convolutional layers (\textit{Transformer-ResNet}) yields a nearly identical result, yet reduces the model size by nearly 45\%. On the other hand, the \textit{Transformer} baseline, which is commonly distributed on public platforms, falls short of the ResNet-inclusive variants in both convergence speed and final bit accuracy.

\subsubsection{Watermark Payload}

As presented in Table~\ref{table:Watermark Payload}, we investigate the impact of varying watermark payload on performance. We adopt a 48-bit payload as the default configuration, which is sufficient to accommodate a vast number of unique user identities and aligns with the standard settings used in prior studies. Our results demonstrate the capacity of AuthenLoRA successfully supporting extended payloads of 80 and 100 bits to address more complex embedding requirements, while maintaining performance degradation within manageable limits.

\begin{table}
	\centering
	\caption{Quantitative comparison of different watermark payloads.}
	\label{table:Watermark Payload}
	\begin{tabular}{lllll} 
	\toprule
	\multicolumn{1}{c}{Watermark Payload} & \multicolumn{1}{c}{BitAcc} & \multicolumn{1}{c}{TPR} & BitAcc(Adv.) & TPR(Adv.)  \\ 
	\cmidrule{1-5}
	48 bits                               & 0.9885                     & 1                       & 0.9525       & 0.9622     \\
	80 bits                               & 0.9699                     & 1                       & 0.9069       & 0.9238     \\
	100 bits                              & 0.9415                     & 0.9980                  & 0.8837       & 0.8805      \\
	\bottomrule
	\end{tabular}
\end{table}

\subsubsection{Watermark Trade-off}
A critical aspect of any watermarking method is its "cost," which for AuthenLoRA primarily relates to the additional model parameters and file size introduced. This cost is directly correlated with the chosen LoRA rank ($R$). While our default configuration employs a rank of $R=320$ to ensure a fair and robust comparison with AquaLoRA, this results in a substantial file size (413.25 MB), which can impede practical model transmission and distribution.

To evaluate the trade-off between model size and watermark efficacy, we assessed performance at reduced ranks. As detailed in Table~\ref{table:Cost}, we observed that even when the LoRA rank is decreased to $R=64$, the file size is significantly reduced to 82.75 MB. This size is far more practical and aligns closely with the typical file sizes of stylized LoRA models found on online platforms. Crucially, even at this highly compressed rank, AuthenLoRA maintains a bit accuracy of $94.84\%$ and a perfect TPR. This success is attributed to the introduction of the extended fine-tuning scope, concurrently proving the strong practicality and efficiency of our method and demonstrating its capability to deliver robust watermarking without imposing prohibitive storage costs.

\begin{table}
	\centering
	\caption{Evaluation of the trade-offs between LoRA rank, model size, and watermark performance.}
	\label{table:Cost}
	\begin{tabular}{llllll} 
	\toprule
	\multicolumn{1}{c}{LoRA Rank} & \multicolumn{1}{c}{BitAcc} & \multicolumn{1}{c}{TPR} & \multicolumn{1}{c}{File Size} & Params     & Cost       \\ 
	\midrule
	320                           & 0.9885                     & 1                       & 413.25 MB                     & 163.51 M & -          \\
	160                           & 0.9792                     & 1                       & 206.69 MB                     & 81.76 M  & 206.56 MB  \\
	80                            & 0.9590                     & 1                     & 103.40 MB                     & 40.86 M  & 309.85 MB  \\
	64                            & 0.9484                     & 1                       & 82.75 MB                      & 32.70 M  & 330.5 MB   \\
	32                            & 0.5827                     & 0                       & 41.43 MB                      & 16.35 M  & 371.82 MB  \\
	\bottomrule
	\end{tabular}
\end{table}

\section{Conclusion}

In this work, we propose \texttt{AuthenLoRA}, a novel watermarking method that effectively protects the copyright of LoRA models by embedding imperceptible watermarks during the generation of stylized images. We jointly optimize style transfer and watermark embedding to achieve tight coupling between the two objectives. A lightweight mapper is introduced to modulate LoRA parameters, enabling dynamic modification of embedded watermarks without retraining. To suppress perceptual watermark residual and reduce false positives, we design the Perceptual Residual Tone Suppression (PRTS) loss and the Zero-Message Regularization strategy. Furthermore, we incorporate ResNet blocks into the fine-tuning scope, significantly improving training efficiency and watermark effectiveness. The limitation of AuthenLoRA lies in the relatively visible watermark residual compared to post-hoc watermarking methods, as well as the increased file size compared to typical LoRA models shared on online platforms. Future work may explore more advanced encoder–decoder architectures to reduce watermark residual, as well as finer-grained fine-tuning scopes and rank configurations to better balance model size and watermark performance. We hope our work contributes to the advancement of digital asset copyright protection.

\bibliographystyle{IEEEtran}
\bibliography{AuthenLoRA_TMM.bib}

\newpage

\appendix

\section{Method}

\subsection{Pre-training for the Watermark Pattern}

\subsubsection{Training Detail}

We employ the AdamW optimizer with a learning rate of $1 \times 10^{-3}$ and a weight decay of $1 \times 10^{-4}$. The model is trained for 50 epochs. The loss function is initially composed with hyperparameters $\lambda = 5$ and $\mu = 0.5$. At epoch 40, we incorporate PRTS Loss and Zero-Message Regularization. The corresponding hyperparameters are set to $\nu = 0.1$ and $p = 0.2$, respectively.

\subsubsection{Pseudo Code}

Algorithm~\ref{alg:pretrain the codec} explains in detail the process of jointly training the watermark encoder and decoder.

\begin{algorithm}
	\caption{Pre-training for the Watermark Pattern}
	\label{alg:pretrain the codec}
	\textbf{Input}: Cover image: $I_o \in \mathbb{R}^{W \times H \times 3}$, Randomly generated watermark information: $m \in \left\{0,1 \right\}^k$\\
	\textbf{Model}: VAE Encoder $E_i$, VAE Decoder $D_i$, Watermark Encoder $E_s$, Watermark Decoder $D_s$, Distortion Layer $\mathcal{T}$\\ 
	\begin{algorithmic}[1] 
		\STATE $Z_o \gets E_i(I_o)$ {//Encode cover image into latent space}
		\STATE Sample $p \sim \text{Uniform}(0,1)$
		\IF{$p < p_{\text{zero}}$}
		\STATE $m \gets \mathbf{0} \in \{0,1\}^k$
		\STATE $Z_w \gets Z_o$ {//No watermark applied}
		\ELSE
		\STATE $Z_s \gets E_s(m)$ {//Encode watermark message}
		\STATE $Z_w \gets Z_o + Z_s$ {//Apply latent watermark}
		\ENDIF
		\STATE $I_w \gets D_i(Z_w)$ {//Decode watermarked latent}
		\STATE $I_r \gets D_i(Z_o)$ {//Decode clean latent}
		\STATE $\tilde{I}_w \gets \mathcal{T}(I_w)$ {//Apply distortion operations}
		\STATE $s^{'} \gets D_s(\tilde{I}_w)$ {//Extract watermark message}
		
		\STATE $\mathcal{L}_{msg} \gets \text{BCE}(m^{'}, m)$
		\STATE $\mathcal{L}_{LPIPS} \gets \text{LPIPS}(I_r, I_w)$
		\STATE $\mathcal{L}_{PRVL} \gets \text{PRVL}(I_r, I_w)$
		\STATE $\mathcal{L}_{PRTS} \gets \text{PRTS}(I_r, I_w)$
		\STATE $\mathcal{L}_{Total} \gets \mathcal{L}_{msg} + \lambda \mathcal{L}_{LPIPS} + \mu \mathcal{L}_{PRVL} + \nu \mathcal{L}_{PRTS}$
		
		\STATE Update $E_s$ and $D_s$ using gradients from $\mathcal{L}_{Total}$
	\end{algorithmic}
\end{algorithm}

\subsection{Stylization and Watermark Embedding}

\subsubsection{Training Detail}

For the LoRA modules, we adopt a rank of 320 and follow a training strategy inspired by \texttt{Kohya\_ss}\cite{KohyaGithub} and \texttt{LyCORIS}\cite{LyCORISGithub}. The optimizer remains AdamW, with a learning rate reduced to $1 \times 10^{-4}$, and training is performed for 10 epochs. During the early phase of training, the total loss is dominated by the watermark loss term, with $\lambda_{wm} = 1$ and the style reconstruction weight $\lambda_{style} = 0$. Once the watermark bit accuracy surpasses a threshold of 0.75 for three consecutive evaluations, $\lambda_{style}$ is gradually elevated to 1, allowing the network to jointly optimize both stylization and watermark embedding objectives.

\subsubsection{Pseudo Code}
Algorithm~\ref{alg:train LoRA and mapper} describes in detail the training process of AuthenLoRA and its watermark mapper.

\begin{algorithm}[H]
	\caption{Stylization and Watermark Embedding} 
	\label{alg:train LoRA and mapper} 
	\textbf{Input}: Cover image: $I_o \in \mathbb{R}^{W \times H \times 3}$, Randomly generated watermark information: $m \in \left\{0,1 \right\}^k$\\ 
	\textbf{Model}: VAE Encoder $E_i$, Watermark Encoder $E_s$, LoRA with parameters $\{\mathbf{A}, \mathbf{B}\}$, Mapper $f_\phi$, Base Diffusion Model $\mathcal{D}_\theta$
	\begin{algorithmic}[1]
		\STATE $Z_o \gets E_i(I_o)$ {//Encode cover image into latent space}
		\STATE $Z_s \gets E_s(m)$ {//Encode message into latent space}
		\STATE $Z_w \gets Z_o + Z_s$ {//Inject watermark into latent space}
		\STATE $\mathbf{v} \gets f_\phi(m)$ {//Mapper generates vector from message}
		\STATE $\mathbf{S} \gets \text{diag}(\mathbf{v})$ {//Convert to diagonal matrix}
		\STATE $\Delta \mathbf{W} \gets \mathbf{A} \times \mathbf{S} \times \mathbf{B}$
		\STATE Sample $\boldsymbol{\epsilon} \sim \mathcal{N}(0, \sigma^2 I)$
		\STATE $\mathbf{Z}_{\text{pred\_wm}} \gets \mathcal{D}_{\theta + \Delta \mathbf{W}}(Z_w + \boldsymbol{\epsilon})$
		\STATE $\mathbf{Z}_{\text{pred\_clean}} \gets \mathcal{D}_\theta(Z_o + \boldsymbol{\epsilon})$
		\STATE $\mathcal{L}_{style} \gets \| \mathbf{Z}_{\text{pred\_wm}} - \boldsymbol{\epsilon}  \|^2$ {//Style loss}
		\STATE $\mathcal{L}_{wm} \gets \| \mathbf{Z}_{\text{pred\_wm}} - \mathbf{Z}_{\text{pred\_clean}} \|^2$ {//Watermark loss}
		\IF{early training phase}
		\STATE $\lambda_{wm} \gg \lambda_{style}$
		\ELSIF{bitwise decoding accuracy $> \tau$}
		\STATE Gradually increase $\lambda_{style}$
		\ENDIF
		\STATE $\mathcal{L}_{total} \gets \lambda_{style} \cdot \mathcal{L}_{style} + \lambda_{wm} \cdot \mathcal{L}_{wm}$
		\STATE Update $\{\mathbf{A}, \mathbf{B}, \phi\}$ via backpropagation on $\mathcal{L}_{total}$
	\end{algorithmic}
\end{algorithm}

\section{Experiments}

\subsection{Experiment Setup}

To comprehensively evaluate the proposed watermark-embedded stylization LoRA framework, we conduct extensive experiments across a variety of stylization types and base models. This section outlines the detailed experimental setup, including model selection, dataset preparation and detailed training configurations.

\IEEEpubidadjcol

\subsubsection{Datasets}
For the \textit{pre-training for the watermark pattern} of the watermark encoder and decoder, we utilize the COCO2017 dataset, from which 10,000 images are randomly sampled for training purposes. This phase ensures that the watermark embedding and extraction network learns to generalize over diverse visual features. For the \textit{stylization and watermark embedding}, we construct four dedicated stylization datasets, each tailored to one of the aforementioned stylization categories. These datasets consist of approximately 1,500 to 2,000 manually curated and annotated images, ensuring sufficient coverage of intra-domain variance.

\subsubsection{Base Models}
Given that different base models inherently exhibit stylistic biases, we select six distinct pre-trained base models as the foundations for stylization and watermark embedding. All selected base models conform to the Stable Diffusion v1.5 architecture to ensure consistency in architecture and parameter compatibility. During training and evaluation, the LoRA scaling factor is uniformly set to 1.0.

\subsubsection{Computing Infrastructure}
All models are implemented using PyTorch framework with version 2.1.2. Training and performance evaluations are conducted on an NVIDIA Titan RTX GPU with 24GB of memory, running Ubuntu 16.04.7 LTS. In addition, we verify the compatibility of our training pipeline on a consumer-grade NVIDIA GeForce RTX 3080 Laptop GPU with 16GB of memory under Windows 10, demonstrating the reproducibility and hardware flexibility of our implementation.

\subsection{Watermark Baselines}
To evaluate the effectiveness of our proposed AuthenLoRA, we compare it against six representative watermarking methods applied at different stages of the image generation process. The Original Dataset baseline involves training a LoRA model solely for stylization using the unmodified dataset, and generating 500 images with 100 prompts randomly generated by GPT-4o, which are then compared with 500 randomly selected samples from the original dataset. This serves as a reference point for visual quality in subsequent evaluations. For post-processing-based watermarking methods and Stable Signature, watermarks are embedded into the 500 generated images and subsequently evaluated.

We do not apply post-processing watermarks directly to the original dataset, as AuthenLoRA is designed to learn and synthesize images that are semantically aligned with the dataset but does not guarantee pixel-level reconstruction. Embedding watermarks in the original dataset and comparing visual fidelity would therefore result in an unfair evaluation. This limitation is also evident in the relatively higher FID scores of images generated by AuthenLoRA when compared to the original dataset. To ensure consistency across evaluation settings, all watermarking methods use the same 500 generated images as the carrier set.

For the Tree-Ring method, we first merge the stylized LoRA model with the base model, and then employ the official pipeline and implementation for both image generation and watermark extraction. In the case of AquaLoRA, we similarly merge the watermarked LoRA and the stylized LoRA into the base model. Image generation is performed via Stable Diffusion, and watermark decoding is conducted using the officially provided decoder.

\subsection{Evaluation Matrix}
We fully evaluate different watermarked images using a well-designed evaluation matrix that includes fidelity, bit accuracy (BitAcc) and true positive rate (TPR).

For Fidelity, we use CLIP\cite{CLIP} to compute the correlation between the text prompts used in generation process and the final image. We use Frechet Inception Distance (FID) \cite{FID} to compare the degree of similarity between the generated image and the original concept. We also use Natural Image Quality Evaluator (NIQE) \cite{NIQE} and Perception based Image Quality Evaluator (PIQE) \cite{PIQE} to evaluate the image visual quality.

For watermark extraction accuracy, we evaluate it by bit accuracy (BitAcc) and true positive rate (TPR). We assess both metrics under two conditions: without any attack and after applying common image processing attacks (Adv.). BitAcc measures the ratio of correctly recovered bits to the total number of embedded bits, reflecting the fidelity of the extracted watermark information. TPR is evaluated under the constraint of a fixed false positive rate (FPR) at $10^{-6}$. 

\subsection{Robustness}

To comprehensively evaluate the robustness of the proposed AuthenLoRA watermarking method, we conducted extensive experiments across diverse scenarios and attacks. 

We first add watermarks to clean images using different watermarking methods. Next, watermarked images are subjected to various common image processing operations to simulate both unintentional degradation (e.g., during online transmission) and deliberate tampering by potential adversaries. The applied distortions include Gaussian blur, color jitter, random cropping, JPEG compression, additive noise, sharpening, and two forms of model-based degradation (SD1.5 and SD2.1), where images are processed using different versions of Stable Diffusion itself via a noise-and-denoise pipeline-a realistic scenario to mimic downstream AI-based reprocessing that may potentially disrupt watermark integrity.

Table~\ref{table:Robustness} summarizes the comparative performance of AuthenLoRA against several watermarking baselines in terms of bit accuracy and TPR. It is evident that AuthenLoRA consistently achieves superior robustness across almost all types of attacks. In particular, under severe distortions such as JPEG compression, additive noise, and the challenging SD1.5 and SD2.1 degradations, AuthenLoRA significantly outperforms existing methods. We evaluate robustness on all four stylized datasets; Table~\ref{table:Robustness} reports the average performance, while Table~\ref{table:Robustness_Detailed} presents the results on each dataset individually.

\begin{table*}[]
	\centering
	\caption{Bit accuracy and TPR of different watermarking methods under various image attacks. The best results under each attack are marked in bold.}
	\begin{tabular}{lllllllllll}
		\hline
		\multicolumn{1}{c}{Method}          & \multicolumn{1}{c}{Clear} & \multicolumn{1}{c}{Blur} & \multicolumn{1}{c}{ColorJitter} & \multicolumn{1}{c}{Crop} & \multicolumn{1}{c}{JPEG} & \multicolumn{1}{c}{Noise} & \multicolumn{1}{c}{Sharpness} & \multicolumn{1}{c}{SD1.5} & \multicolumn{1}{c}{SD2.1} & \multicolumn{1}{c}{Avg} \\ \hline
		\textit{Bit Accuracy(\%)$\uparrow$} &                           &                          &                                 &                          &                          &                           &                               &                           &                           &                         \\
		DWT-DCT                             & 0.8553                    & 0.6928                   & 0.7338                          & 0.5008                   & 0.4919                   & 0.4495                    & 0.5891                        & 0.4851                    & 0.4856                    & 0.5536                  \\
		DWT-DCT-SVD                         & 0.9794                    & 0.9614                   & 0.5647                          & 0.5125                   & 0.6830                   & 0.5453                    & 0.8479                        & 0.5520                    & 0.5083                    & 0.6469                  \\
		RivaGAN                             & \textbf{0.9927}           & \textbf{0.9914}          & 0.9715                          & \textbf{0.9912}          & 0.9441                   & 0.6708                    & \textbf{0.9892}               & 0.5720                    & 0.5274                    & 0.8322                  \\
		Tree-Ring                           & 0.9914                    & 0.7303                   & 0.9430                          & 0.9631                   & 0.7992                   & 0.7835                    & 0.9721                        & 0.4734                    & 0.4506                    & 0.7837                  \\
		Stable Signature                    & 0.9359                    & 0.9362                   & 0.9166                          & 0.8817                   & 0.9242                   & 0.9119                    & 0.9329                        & 0.8478                    & 0.7987                    & 0.8937                  \\
		AuthenLoRA                          & 0.9885                    & 0.9887                   & \textbf{0.9833}                 & 0.9058                   & \textbf{0.9816}          & \textbf{0.9752}           & 0.9874                        & \textbf{0.9342}           & \textbf{0.8639}           & \textbf{0.9525}         \\ \hline
		\textit{TPR@FPR=$10^{-6}\uparrow$}  &                           &                          &                                 &                          &                          &                           &                               &                           &                           &                         \\
		DWT-DCT                             & 0.671                     & 0.305                    & 0.4915                          & 0                        & 0                        & 0                         & 0.2395                        & 0                         & 0                         & 0.1295                  \\
		DWT-DCT-SVD                         & 0.9625                    & 0.926                    & 0.2945                          & 0                        & 0.2025                   & 0                         & 0.6455                        & 0.0105                    & 0                         & 0.2599                  \\
		RivaGAN                             & 0.9935                    & 0.9925                   & 0.946                           & 0.99                     & 0.904                    & 0.1645                    & 0.9895                        & 0.0035                    & 0                         & 0.6238                  \\
		Tree-Ring                           & 0.944                     & 0.905                    & 0.8248                          & 0.6391                   & 0.6177                   & 0.5822                    & 0.6200                        & 0.4446                    & 0.3338                    & 0.6209                  \\
		Stable Signature                    & \textbf{1}                & 0.248                    & 0.973                           & \textbf{0.996}           & 0.46                     & 0.444                     & 0.993                         & 0                         & 0                         & 0.5721                  \\
		AquaLoRA                            & 0.914                     & 0.9155                   & 0.8805                          & 0.805                    & 0.8955                   & 0.8745                    & 0.9065                        & 0.686                     & 0.485                     & 0.8061                  \\
		AuthenLoRA                          & \textbf{1}                & \textbf{1}               & \textbf{0.9995}                 & 0.9325                   & \textbf{1}               & \textbf{0.9985}           & \textbf{0.9995}               & \textbf{0.986}            & \textbf{0.7815}           & \textbf{0.9622}         \\ \hline
	\end{tabular}
	\label{table:Robustness}
\end{table*}

\subsection{Zero-Message Regularization}

To simulate a realistic environment and quantitatively evaluate the reliability of different watermark decoders, we conduct a binary classification experiment using 500 images with embedded watermarks and 500 clean (non-watermarked) images. For each method, we measure Precision, Recall, and false positive rate (FPR), based on whether the decoder correctly identifies the presence or absence of a watermark. A predicted watermark is considered valid if the decoded message is not an all-zero or all-one vector; otherwise, it is interpreted as recognizing a clean image.

Precision reflects the proportion of correctly predicted watermarked images among all images that were predicted to contain a watermark. A high precision value indicates strong reliability of positive detections and ensures that false claims of ownership (false positives) are minimized. Recall measures the proportion of actual watermarked images that are correctly identified, directly reflecting the decoder's ability to recover watermarks when they are indeed present.

Most critically in our scenario, the false positive rate (FPR) quantifies the proportion of clean, non-watermarked images that are incorrectly identified as containing a watermark. A high FPR poses a severe risk in real-world copyright enforcement, as it may lead to false accusations or wrongful copyright claims over unrelated content. Therefore, reducing FPR is of paramount importance.

As shown in Table~\ref{table:Zero-Message Regularization}, post-process methods achieve relatively high recall, but suffer from poor precision and high FPR, indicating that they frequently produce hallucinated watermarks on clean images. During-diffusion methods, including Tree-Ring, Stable Signature, and AquaLoRA, also exhibit perfect recall but extremely poor precision and FPR of 1.0, revealing that they misclassify all clean images as watermarked and thus cannot reliably reject false positives.

In contrast, our method with zero-message regularization achieves perfect recall, showing that the model is able to successfully recognize all images with watermarks. Meanwhile, our method improves the precision significantly to 0.9887 and reduces the FPR to 0.0115. This result clearly demonstrates the benefit of zero-message regularization in suppressing spurious watermark decoding on non-watermarked content. Without zero-message regularization mechanism, the decoder falls back to the baseline behavior of always predicting a watermark, resulting in poor FPR and precision.

\begin{table*}[]
	\centering
	\caption{Comparison of the performance of our proposed method with other watermarking methods based on post-processing and during the diffusion process. We control the FPR at $10^{-6}$ and calculate the TPR. Since Tree-Ring is a zero-bit watermark method, the bit accuracy cannot be calculated, and we only calculated the TPR. The best results are presented in bold text.}
\begin{tabular}{llcccccccc} 
		\toprule
		\multicolumn{1}{c}{\multirow{2}{*}{Method}} & \multicolumn{1}{c}{\multirow{2}{*}{Capacity}} & \multicolumn{4}{c}{Fidelity}                                                                                                                           & \multicolumn{1}{c}{\multirow{2}{*}{BitAcc}} & \multicolumn{1}{c}{\multirow{2}{*}{TPR}} & \multicolumn{1}{c}{\multirow{2}{*}{BitAcc(Adv.)}} & \multicolumn{1}{c}{\multirow{2}{*}{TPR(Adv.)}}  \\ 
		\cmidrule{3-6}
		\multicolumn{1}{c}{}                        & \multicolumn{1}{c}{}                          & \multicolumn{1}{c}{CLIP$\uparrow$} & \multicolumn{1}{c}{FID$\downarrow$} & \multicolumn{1}{c}{NIQE$\downarrow$} & \multicolumn{1}{c}{PIQE$\downarrow$} & \multicolumn{1}{c}{}                        & \multicolumn{1}{c}{}                     & \multicolumn{1}{c}{}                              & \multicolumn{1}{c}{}                            \\ 
		\midrule
		\multicolumn{10}{l}{\textbf{Anime Character}}                                                                                                                                                                                                                                                                                                                                                                                                       \\
		\multicolumn{10}{l}{Post-Process}                                                                                                                                                                                                                                                                                                                                                                                                                   \\
\bottomrule
	\end{tabular}
	\label{tabel:Overall Comparison in Fidelity and Accuracy}
\end{table*}

\begin{table*}[]
	\centering
	\caption{Bit accuracy and TPR of different watermarking methods under various image attacks.}
	\begin{tabular}{lcccccccccc} 
		\toprule
		\multicolumn{1}{c}{Method}          & Clear                & Blur                 & ColorJitter          & Crop                 & JPEG                 & Noise                & Sharpness            & SD1.5                & SD2.1                & Avg                   \\ 
		\midrule
		\multicolumn{11}{l}{\textbf{Anime Character}}                                                                                                                                                                                                                              \\
		\textit{Bit Accuracy(\%)$\uparrow$} & \multicolumn{1}{l}{} & \multicolumn{1}{l}{} & \multicolumn{1}{l}{} & \multicolumn{1}{l}{} & \multicolumn{1}{l}{} & \multicolumn{1}{l}{} & \multicolumn{1}{l}{} & \multicolumn{1}{l}{} & \multicolumn{1}{l}{} & \multicolumn{1}{l}{}  \\
		DWT-DCT                             & 0.8920               & 0.6935               & 0.8512               & 0.5047               & 0.4881               & 0.4313               & 0.6218               & 0.4854               & 0.4843               & 0.5700                \\
		DWT-DCT-SVD                         & 1.0000               & 1.0000               & 0.6599               & 0.5012               & 0.6265               & 0.5762               & 0.8999               & 0.6057               & 0.5147               & 0.6730                \\
		RivaGAN                             & 1.0000               & 1.0000               & 0.9981               & 0.9999               & 0.9764               & 0.6566               & 0.9999               & 0.6138               & 0.5274               & 0.8465                \\
		Tree-Ring                           & 0.9883               & 0.6873               & 0.9095               & 0.9335               & 0.7346               & 0.6838               & 0.9547               & 0.4400               & 0.4237               & 0.7209                \\
\bottomrule
	\end{tabular}
	\label{table:Robustness_Detailed}
\end{table*}

\begin{table*}
	\centering
	\caption{Performance comparison of six watermark decoders on a binary classification task, using 500 watermarked and 500 non-watermarked images. The best result is highlighted in bold. }
	\begin{tabular}{lccccccccc} 
		\toprule
		\multicolumn{1}{c}{Method} & \multicolumn{1}{l}{TP} & \multicolumn{1}{l}{TN} & \multicolumn{1}{l}{FP} & \multicolumn{1}{l}{FN} & Precision            & Recall               & \multicolumn{1}{l}{Accuracy} & \multicolumn{1}{l}{F1-Score} & FPR                   \\ 
		\midrule
		\multicolumn{10}{l}{\textbf{Anime Character}}                                                                                                                                                                                                                      \\
		\textit{Post-Process}      & \multicolumn{1}{l}{}   & \multicolumn{1}{l}{}   & \multicolumn{1}{l}{}   & \multicolumn{1}{l}{}   & \multicolumn{1}{l}{} & \multicolumn{1}{l}{} & \multicolumn{1}{l}{}         & \multicolumn{1}{l}{}         & \multicolumn{1}{l}{}  \\
		DWT-DCT                    & 460                    & 221                    & 279                    & 40                     & 0.6225               & 0.9200               & 0.6810                       & 0.7425                       & 0.5580                \\
		DWT-DCT-SVD                & 500                    & 44                     & 456                    & 0                      & 0.5230               & 1.0000               & 0.5440                       & 0.6868                       & 0.9120                \\
		RivaGAN                    & 500                    & 0                      & 500                    & 0                      & 0.5000               & 1.0000               & 0.5000                       & 0.6667                       & 1.0000                \\
		\textit{During-Diffusion}  &                        &                        &                        &                        &                      &                      &                              &                              &                       \\
\end{tabular}
	\label{table:Zero-Message Regularization}
\end{table*}

\end{document}